-----------------------------------------------------------------------
\documentstyle[12pt]{article}
\newcommand{\be}{\begin{equation}}
\newcommand{\ee}{\end{equation}}
\newcommand{\bq}{\begin{eqnarray}}
\newcommand{\eq}{\end{eqnarray}}

\begin{document}

\pagestyle{empty}
\begin{flushright}
{\large UBCTP 92-23\\
revised \\
August 1992}
\end{flushright}
\vspace{0.4cm}
\begin{center}
{\large \bf
AREA PRESERVING DIFFEOMORPHISMS AND  $W_{\infty}$
    SYMMETRY IN A  $2+1$ CHERN-SIMONS THEORY
}\\ \vspace{1 cm}
{\large Ian I. Kogan} \footnote{ On leave of ITEP, Moscow, USSR.},
 \footnote{This work is supported in part by the Natural
Sciences and Engineering Research Council of Canada}\\

\vspace{0.6 cm}
{ Department of Physics, University of British Columbia\\
Vancouver, B.C., Canada V6T1Z1}\\
 \end{center}
\vspace{0.4cm} \noindent
\begin{center}
{\bf Abstract}
\end{center}
 We discuss the $W_{\infty}$ symmetry in the $2+1$ gauge theory with
 the Chern-Simons term. It is shown that the generators of
 this symmetry act on the ground state as the canonical transformations
 in the phase space. We shall also discuss the analogy between
 discrete states in $c=1$ string theory and Landau level
 states in $2+1$
 gauge theory with Chern-Simons term.

\newpage
\pagestyle{plain}
\setcounter{page}{1}
\stepcounter{subsection}
\section{Introduction}

 The  $W_{\infty}$ algebras \cite{w} can be
 obtained after taking  $N\rightarrow\infty$ limit of the Zamolodchikov's
 $W_{N}$ algebras
($W_{2} = V$ is the Virasoro algebra)
 \cite{zam}, which are not Lie algebras at finite $N>2$ due to the non-linear
 terms in the commutation relations. In the
  $N\rightarrow\infty$ limit the structure of the commutators
 is much more simpler \cite{bilal} and, depending on the
  limiting procedure  which is is not uniquely defined,   one
  gets different algebras.  One particular limit \cite{w} is known as
 $w_{\infty}$ algebra
\bq
[w^{(i)}_{m},w^{(j)}_{n}] = \bigl( (j-1)m -(i-1)n\bigr) w^{(i+j-2)}_{m+n}
\label{w}
\eq
where $w^{(i)}_{m}$ is a generator of conformal spin $i$.

It is amusing that the algebra (\ref{w}) is an  algebra of area-preserving
 (or symplectic)  diffeomorphisms of the two-dimensional manifolds,
  for example it is an  infinite-dimensional algebra    of canonical
 transformation  in a two-dimensional phase space $(p,q)$.  It is interesting
 that $w_{\infty}$ algebra generates  the symmetry of the relativistic
 membranes after gauge fixing - this symmetry and it connection with
 the $SU(\infty)$ were considered in \cite{membrane}.

Recently it was found \cite{kpw} that
  $w_{\infty}$ symmetry  is the dynamical symmetry for
 discrete states \cite{discrete}   of the two dimensional strings \footnote
{for $SL(2;R)/U(1)$ cosets the infinite-dimensional
 symmetries were considered in \cite{yuwu}} and area
 preserving diffeomorphisms are the canonical transformations
 acting on the phase space of the inverted harmonic
 oscillator which is arised in the
  $c=1$  matrix models
 \cite{matrix}.

  In this letter we would like to discuss another interesting
 physical system where  area preserving diffeomorphisms and
 $w_{\infty}$ symmetry are arising as  canonical transformations -
 this is topologically massive gauge theory \cite{tmgt}, i.e.
 $2+1$-dimensional gauge theory  with the Chern-Simons term.
  It  was shown  \cite{landau}  that the  Hilbert space
 of the theory is a direct  product of the massive gauge particles
 Hilbert space (one free massive particle in the most simple
 $U(1)$ case)
 and  some quantum-mechanical Hilbert space. In the
  $U(1)$ case this QM Hilbert space  is the product of the
  $g$ copies (for a genus $g$ Riemann surface) of the
 Hilbert space for the  Landau problem on the torus.
  In the infinite mass limit
  all levels except the first one are  decoupled  as well
 as massive particles Hilbert space and we have only
 the first Landau level. It is easy to see
 that the first Landau level becomes the  phase space for
 the pure Chern-Simons theory and canonical transformations
 acting on the  phase space are nothing but area preserving
 diffeomorphisms.

 The organization of the paper is as follows. In section
 2, which bears essentially a review character, we
 consider the canonical phase space of the
 topologically massive gauge theory and describe
 canonical quantization and Landau levels picture.
 We shall also  discuss analogy between discrete
 states in $c=1$ string theory and  quantum mechanical
 states in gauge theory.
 In section 3 the area-preserving diffeomorphisms
  and $w_{\infty}$ algebra   will be considered.

  \section{Canonical quantization of the $2+1$
 TMGT}

Let us consider  the most simple case of the abelian gauge theory with
  the  action \cite{tmgt}:
\begin{equation}
S_{U(1)} =
 -\frac{1}{4\gamma}\int \sqrt{-g}g^{\mu\alpha}g^{\nu\beta}F_{\mu\nu}
 F_{\alpha\beta} +
\frac{k}{8\pi}\int \epsilon_{\mu\nu\lambda}
A_{\mu}\partial_{\nu}A_{\lambda}
\label{eq:tmgt}
\end{equation}
To perform canonical quantization one choses the
  $A_{0}=0$ gauge.
 One  can represent vector-potential on the plane as:
\begin{equation}
A_{i} = \partial_{i}\xi + \epsilon_{ij}\partial_{j}\chi \label{eq:decomp1}
\end{equation}
Substituting this decomposition into constraint
\begin{equation}
\frac{1}{\gamma}\partial_{i}\dot{A}_{i} + \frac{k}{4\pi} \epsilon_{ij}F_{ij} =
0, \label{eq:constraint}
\end{equation}
one  gets
 $\partial^{2} \dot{\xi} = (k\gamma/2\pi)\partial^{2} \chi $.
Neglecting all possible zero modes we  put $\dot{\xi} =
 (k\gamma/2\pi) \chi = (M/2)\chi$. Substituting this constraint into
action (\ref{eq:tmgt}) one gets the   action
\begin{eqnarray}
S = \frac{1}{2\gamma} \int(\partial_{i}\dot{\chi})^{2} -
(\partial^{2}\chi)^{2} - M^{2}\chi\partial^{2}\chi
\end{eqnarray}
which becomes the free massive particle action for the  field $\Phi =
\sqrt{\partial^{2}/ \gamma} \chi$:
\begin{eqnarray}
S = \frac{1}{2} \int \dot{\Phi}^{2} - (\partial_{i}\Phi)^{2} -
M^{2}\Phi^{2}
\end{eqnarray}
To get this free action it was extremely important to use
 the  constraint
(\ref{eq:constraint}). However  there are some
 field configurations which are not affected by this constraint.
 It is easy to see  that on the plane   the
constant fields $A_{i}(x,t) = {\bf A}_{i}(t)$ are not affected
 by the constraint (\ref{eq:constraint}) - because both terms
  $F_{ij}$ and  $\partial_{i}E_{i}$ are zero for
 space-independent vector potential ( but not electric field
 $E_{i} = \dot{A}_{i}$ ).
  For this field one  gets the
quantum mechanical ( no coordinate dependence) Lagrangian:
\begin{equation}
L = \frac{1}{2\gamma}\dot{{\bf A}}_{i}^{2} - \frac{k}{8\pi}\epsilon_{ij}{\bf
A}_{i}\dot{{\bf A}}_{j}  \label{eq:Landau}
\end{equation}
which describes the particle on the plane ${\bf A}_{1},{\bf A}_{2}$ in
magnetic field - the famous  Landau problem.

    Let us consider the configuration and phase spaces of this
 problem. From Lagrangian (\ref{eq:Landau}) one gets the
 canonical momenta:
\bq
{\bf P}_{i} = \frac{\partial L}{\partial{\bf A}_{i}} =
 \frac{1}{\gamma}\dot{{\bf A}}_{i} + \frac{k}{8\pi}\epsilon_{ij}{\bf
A}_{j}
\label{momentum}
\eq
with the usual commutation relations (or Poisson brackets
 in the classical limit $[,] \rightarrow -i\{,\}$)
\bq
[{\bf P}_{i}, {\bf P}_{j}] =  [{\bf A}_{i}, {\bf A}_{j}] = 0;
\;\;\;[{\bf P}_{i}, {\bf A}_{j}] = -i\delta_{ij}
\eq
The canonical Landau Hamiltonian and the eigenvalues $E_{n}$ are
\bq
H = -\frac{\gamma}{2}(\frac{\partial}{\partial{\bf A}_{i}} -
i\frac{k}{8\pi} \epsilon_{ij}{\bf A}_{j})^{2} \;\;\;\;\;\;\;
E_{n} = (n +
1/2) M
\label{eq:landham}
\eq
where the gap equals to the topological mass
  $M = \gamma k/4\pi$.

Let us note that variables $({\bf A}_{1}$ and ${\bf A}_{2})$
 belong to  the configuration
 space, however if reduced to  the first Landau level the
 configuration space is transformed into the phase space. It is
 easy to see that in the limit $\gamma \rightarrow \infty$
 (which corresponds to the reduction on the first Landau level)
 ${\bf P}_{i} =  \frac{k}{8\pi}\epsilon_{ij}{\bf A}_{j}$   and
 one of the coordinates becomes a conjugate momentum for another
 one. Thus   $({\bf A}_{1}, {\bf A}_{2})$   plane is the
 configuration space for the topologically massive gauge theory
 and becomes the phase space for pure Chern-Simons theory which
 is an exactly solvable  $2+1$ dimensional topological field theory
 \cite{witten}
This picture of the Landau levels were considered  in \cite{landau} and it
was proved, that due to the degeneracy on each Landau level one  gets
long-range part of the gauge propagator in TMGT (the propagator in pure CS
theory), in spite of the fact that there are no massless particles in theory.
In other words, the Hilbert space of the pure Chern-Simons
 theory (i.e. in the limit $M\rightarrow\infty$) is the first Landau level.

Is it possible to consider
 a  constant gauge field  as a physical, i.e. gauge invariant
  variable  in the theory ?
  The answer is positive for  2-dimensional Riemann surfaces  different
from a  sphere. It is well-known
 that any  one-form $A$ can be uniquely decomposed  according to
 Hodge
theorem as
\begin{eqnarray}
 A = d\xi + \delta\chi + {\bf A} \nonumber \\
d{\bf A} = \delta{\bf A} = 0
\end{eqnarray}
which generalizes the decomposition (\ref{eq:decomp1}); the harmonic form
${\bf A}$ equals
\begin{eqnarray}
{\bf A} = \sum_{p=1}^{g} ({\bf A}^{p}\alpha_{p} + {\bf B}^{p}\beta_{p})
\label{1forms}
\end{eqnarray}
where $\alpha_{p}$ and $\beta_{p}$ are canonical harmonic 1-forms on the
Riemann surface of  genus $g$. It is well known that there
 are precisely $2g$ harmonic 1-forms  for any genus $g$ Riemann surface
 and  one gets $g$ copies of the Landau problem  for each conjugate
 pair ${\bf A}^{p}, {\bf B}^{p}$. The Hamiltonian for each pair
 is the usual Landau Hamiltonian where ${\bf A}, {\bf B}$ corresponds
 to ${\bf A}_{1}, {\bf B}_{2}$ . There is also some dependence
 on the moduli of the Riemann surface due to   the $F^{2}$  term
 in (\ref{eq:tmgt})   dependence
 on metric $g_{\mu\nu}$. It is easy to see that
 for $g_{00} = 1$ and $g_{ij}= \rho(x) h_{ij} (\tau)$ the
  $F_{0i}^{2}$ term does not depend  on  conformal factor
 $\rho$.  Let us  consider dependence on the moduli $\tau$ in the
 most simple case of a  torus where $\tau$ is a complex number  and
 metric $h_{ij}$ can be parametrized as
\begin{eqnarray}
 h^{ij} = \frac{1}{(Im\tau)^{2}} \left( \begin{array}{cc}
1 &
Re\, \tau\\
 & \\
Re \, \tau & |\tau|^{2} \\ \end{array}\right)  \;\;\;\;\;
h_{ij} = \left( \begin{array}{cc}
|\tau|^{2}&
-Re\, \tau\\
 & \\
-Re \, \tau & 1 \\ \end{array}\right) \label{h}
\end{eqnarray}
 and  $h = det h_{ij} = (Im\tau)^{2}$.
  Lagrangian takes the form
 \begin{equation}
L = \frac{1}{2\gamma}\sqrt{h}h^{ij}\dot{{\bf A}}_{i}
\dot{{\bf A}}_{j}  - \frac{k}{8\pi}\epsilon^{ij}{\bf
A}_{i}\dot{{\bf A}_{j}}  \label{landau}
\end{equation}
which can be transformed to diagonal form  (\ref{eq:Landau})
 for new fields
\bq
{\bf A}_{(a)} = e^{i}_{(a)}{\bf A}_{i}
\eq
 where zweibein $e^{i}_{(a)}$ defines the metric
$h^{ij} = e^{i}_{(a)}e^{j}_{(b)} \delta^{(a)(b)}$
 and  $\epsilon^{(a)(b)} e^{i}_{(a)}e^{j}_{(b)} \sim\epsilon^{ij}$
  It is easy to  find that
\bq
 \left(\begin{array}{c}
 {\bf A}_{(1)}\\{\bf A}_{(2)} \\ \end{array}\right)
  = \left( \begin{array}{cc}
 1 &  Re\, \tau \\
 & \\
0 &  Im\, \tau \\ \end{array}\right)
\left(\begin{array}{c}
 {\bf A}_{1}\\{\bf A}_{2} \\ \end{array}\right)
\label{new}
  \eq
   In terms of the new variables the Lagrangian
 (\ref{landau}) takes the form
 \begin{equation}
L = \frac{1}{2\gamma~ Im \,\tau }\dot{{\bf A}}_{(i)}^{2}
  - \frac{k }{8\pi ~Im \,\tau}\epsilon^{(i)(j)}{\bf
A}_{(i)}\dot{{\bf A}}_{(j)}  \label{diaglandau}
\end{equation}
and we see that  the Chern-Simons coefficient depends on
 moduli: $k \rightarrow k/Im\,\tau$. However the mass gap
 is unchanged  because $\gamma$ is also
 changed $\gamma  \rightarrow\gamma~
 Im\,\tau$
 and  $M = \gamma k/4\pi$ does not depend
 on $\tau$.

 Thus we get the Landay problem on the plane $ ({\bf A}_{(1)},
 {\bf A}_{(2)})$.
   However we forgot about  large  gauge transformations acting
  on the
quantum-mechanical coordinates  ${\bf A}_{i}\rightarrow{\bf A}_{i}+
 2\pi\, N_{i}$,
 where $N_{i}$ are integers. This is due to
 the fact that gauge-invariant objects - Wilson lines
 $W(C) = exp(i\oint_{C} A_{\mu} dx^{\mu})$, are not changed
 under these transformations (we normalize  here  the lengths
   of all the fundamental cycles to one)
 and one can consider torus  $0\leq{\bf A}_{i}<2\pi$
  with the area $(2\pi)^{2}$.   However
 after we  consider the new variables   ${\bf A}_{(i)}$
 one gets the torus (see (\ref{new}))
  generated by the shifts $2\pi$ and $2\pi\tau$
  with  an area $S = (2\pi)^{2}~Im \tau$.

  Let us note that being reduced
  to the first Landau level
  this torus becomes the phase space - thus for the consistent quantization
  this area must be proportional to the  integer
  (the   total number of the states must be integer).
   It is known
 that the density of states $\rho$  on Landau level equals
 to
 $H/2\pi$, where $H$ is a magnetic field. In our case the
 "magnetic field" in   $({\bf A}_{(1)}, {\bf A}_{(2)})$  plane can be
 easily obtained from (\ref{diaglandau}) and equals to
 ${\cal H} = (k/4\pi~ Im\,\tau)$, thus the  total number of states
 will be $N = (1/2\pi) (k/4\pi~ Im\,\tau) \times (2\pi)^{2}~Im\,\tau =
 k/2$.  and does not depend on $\tau$  but only on $k$.
  One can factorize over whole large gauge transformations only for even
 $k$, for general $k = 2m/n$ only the shifts
 ${\bf A}_{i}\rightarrow{\bf A}_{i}+n N_{i}$,
  are compatible
 with the dynamics and one gets $0\leq{\bf A}_{i}
 < 2\pi n$ so the total area is now $(2\pi n)^{2}$ and the number of states
 will be $2mn$.

In the case of the genus $g$ Riemann surface one gets
 $g$ conjugate pairs
  ${\bf A}^{p}, {\bf B}^{p}$, $p = 1,\cdots, g$.   After diagonalization
  one finds that there are $g$ copies of the Landau problem and
  the  total Hilbert space of the abelian topologically massive gauge
theory
\bq
{\Large H} = {\Large H}_{\Phi}\otimes\prod_{i=1}^{g}{\Large H}_{{\bf A}}
\eq
is the product of the free massive particles Hilbert space and $g$ copies
of the  Landau problem's  Hilbert space. One can consider the non-abelian
 theories in a similar way.

 Let us also note some analogy between  gauge fields (\ref{1forms}) and
 discrete states in string theory \cite{discrete}. The
 physical states in the string theory must satisfy the Virasoro
 constraints modulo gauge transformations - in a
 complete analogy with the  constraint (\ref{eq:constraint}) in
 a gauge theory.   As was stressed
 by Polyakov in \cite{discrete} the discrete states in $c=1$ string
 theory exists in the string theory only because Virasoro constraints
 $L_{n}|\psi> = 0,~~n>0$ do not act effectively on some excited
 states, contrary to the generic case when the only non-trivial
 solutions of the Virasoro constraints are gauge artifacts.
 The simplest example is provided by the level one
 operator in open string theory (we are  using here Klebanov
 and Polyakov \cite{kpw} notation)  $ e_{\mu}(p,\epsilon)
 \partial X^{\mu} exp(ipX + \epsilon \phi)$ where $\phi$
 and $X$ are the Liouville and matter fields and  $X^{\mu} =
 (\phi, X)$. The conditions that this operator is physical
  are the follows:
\bq
 (f_{\mu} + b_{\mu}) e^{\mu}(f) = 0;~~~f_{\mu}(f^{\mu}
 + b^{\mu}) = 0
\eq
 where $f_{\mu} = (\epsilon, p)$, $b_{\mu} = (2,0)$ and
    the signature is $(+,-)$.  For generic
 $p$ the only solution of this constraints is the pure gauge
 $e_{\mu}(f) \sim f_{\mu}$. However for
 $f_{\mu} = 0$ or $-b_{\mu}$ one constraint
 disappears and it is possible to show that one can
 gauged away only $e_{0}$ component, but not $e_{1}$. In the same
 way all other states at higher levels can be obtained.

 It is evident
 that  this picture is extremely similar  to the one which was
 considered here - when the quantum-mechanical degrees of freedom
  (\ref{1forms})
 were not affected by the gauge constraint (\ref{eq:constraint})
 and becomes the new physical degrees of freedom. Thus we have
 very intriguing analogy between $c=1$ strings with a $1+1$
 dimensional
 physical field (massless "tachyon") and discrete states
 and topologically massive gauge theory with a
 $2+1$ dimensional  physical field (topologically massive
 "photon") and quantum-mechanical degrees of freedom (states
 on the Landau levels).  What is extremely interesting
 in this analogy is the $w_{\infty}$ symmetry which
 exists for both discrete states  \cite{kpw},\cite{matrix}
 and states on Landau level as we shall  see in the
 next section.

\section{Canonical transformation on the Landau level
 and $w_{\infty}$ algebra}

 Let us consider  the Landau  Hamiltonian
\bq
 H = \frac{\gamma}{2}({\bf P}_{i} -
 \frac{k}{8\pi} \epsilon_{ij}{\bf A}_{j})^{2}
\label{H}
\eq
 where ${\bf A}_{i}$ and ${\bf P}_{j}$ are the canonical coordinates
 in the four-dimensional phase space.  In order to study the
 symmetries of the problem in more details let us remember
 some well-known facts about canonical transformations (see,
 for example \cite{arnland}). By
 definition canonical transformations are
   diffeomorphisms
 of the phase space which preserve the symplectic structure
 $\omega = \sum_{p} dq^{l}\wedge dp^{l}$. It is possible
 to show that conservation of the symplectic structure leads
 to the Liouville theorem about the conservation of the phase
 space volume during the time.

The canonical transformations are usually  defined by the
 generation function depending   on both  old
 ( $p$ or $q$) and new ($P$ and $Q$) phase space
 coordinates, for example one can consider arbitrary
 $F(q,Q)$ and put
 \bq  p_{i} = \partial F/\partial q_{i};~~~
 P_{i} = -\partial F/\partial Q_{i}
\eq
 It is easy to see that $P,Q$ are new canonical coordinates.
There is however another representation, namely one can consider
 evolution with respect to some "Hamiltonian" $\Phi(p,q)$ (which is arbitrary
 function on the phase space and has  nothing common with the
 physical Hamiltonian). The change
 in quantities $p$ and $q$ during this evolution may itself be
 regarded as a series of canonical transformations.  Let
 $p$ and $q$ be the values of the canonical variables at time $t$
 and $P$ and $Q$ are their values at another time $t + \tau$.
 The latter are some function of the former, depending on $\tau$
  as on parameter
\bq
Q = Q(q,p;\tau),~~~~~~P=P(q,p;\tau)
\eq
These formulae can be considered as the canonical transformation
from the old coordinates $p,q$ to the new ones $P,Q$. This
 representation is convenient for the infinitesimal transformation,
 when $\tau \rightarrow 0$. In this case using Hamiltonian
 equations of motion with "Hamiltonian" $\Phi(p,q)/\tau$ one gets
\bq
 Q_{i} = q_{i} + \dot{q}_{i}\tau = q_{i} + \{q_{i}, \Phi\}; ~~~~~
 P_{i} = p_{i} + \dot{p}_{i}\tau = p_{i} + \{p_{i}, \Phi\}
\label{can1}
\eq
 where
\bq
\{A, B \} = \sum_{i} \frac{\partial A}{\partial q_{i}}
\frac{\partial B}{\partial p_{i}} -\frac{\partial B}{\partial q_{i}}
\frac{\partial A}{\partial p_{i}}
\label{brackets1}
 \eq
 is the usual Poisson brackets.

 Let us now consider canonical transformations  acting
 on the Landau problem phase space. The general canonical
 transformations are acting on the whole four-dimensional
 phase space  and after quantization they will mix different
 Landau levels.  However there is a  special subgroup of
 the canonical transformations acting  on the two-dimensional
 subspace of the phase space commuting with the Hamiltonian.
 This means that this transformations do not mix different
 Landau levels and thus acting on each Landau level as on
 two-dimensional phase space. To define this group let us
 notice  that  Hamiltonian  (\ref{H})  depends
 on two variables $a$ and $a^{+}$
\bq  a^{+} = 2{\bf P}_{\bar{z}} +
 i \frac{k}{8\pi} {\bf A}_{z}; ~~~
a = 2{\bf P}_{z} -
 i \frac{k}{8\pi} {\bf A}_{\bar{z}};~~~[a,a^{+}] =
- i\{a,a^{+}\} =  \frac{k}{2\pi}
\label{a}
\eq
 and
\bq
 H = \frac{\gamma}{4} \bigl( aa^{+} + a^{+}a \bigr)
\eq
  Here  ${\bf A}_{z,\;\bar{z}} = {\bf A}_{1}\pm i {\bf A}_{2}$ and
${\bf P}_{z,\; \bar{z}}= -i\partial/\partial {\bf A}_{z,\;\bar{z}}$
 are the corresponding  conjugate momenta.  There is
 another pair  $b$ and $b^{+}$ commuting with $a$ and $a^{+}$
\bq
 b^{+} =2{\bf P}_{z} +
 i \frac{k}{8\pi} {\bf A}_{\bar{z}};~~~ b =  2{\bf P}_{\bar{z}} -
 i \frac{k}{8\pi} {\bf A}_{z}; ~~~
 [b,b^{+}] =
- i\{b,b^{+}\} =  \frac{k}{2\pi}
\label{definitionb}
\eq
 Now what means the restriction on the first level - this is
  to the pure Chern-Simons theory, i.e. taking the limit
 $\gamma \rightarrow \infty$. In this limit one gets from
  (\ref{momentum})
\bq
2{\bf P}_{\bar{z}} = {\bf P}_{1} + i {\bf P}_{2} =
  -i\frac{k}{8\pi}{\bf A}_{z};~~2{\bf P}_{z} = {\bf P}_{1} - i {\bf P}_{2} =
  i\frac{k}{8\pi}{\bf A}_{\bar{z}}
 \eq
and then
\bq
 a = a^{+} = 0; ~~~~ b^{+} =i \frac{k}{4\pi} {\bf A}_{\bar{z}};~~
 b =  -
 i \frac{k}{4\pi} {\bf A}_{z}
\eq
 The physical  meaning of  this reduction is the follows -
 operators  $a$ and $a^{+}$
 acting on the state at the given level $n$ shift it to $n\mp 1$.
 To be at a given Landau level  we  must put these operators
 to zero after which   $b$ and $b^{+}$
 play the role of the coordinate on the reduced  phase space
 (which is the  total phase space in the pure Chern-Simons
 theory)  and
  one  can consider the canonical transformations  on this space.

 As we know from   (\ref{can1}) and (\ref{brackets1}) the canonical
 transformations acting on the two-dimensional phase
 space $(q,p)$  are defined by
\bq
 \delta q =  \{q, W(p,q)\}=\frac{\partial W(p,q)}{\partial p}
 ; ~~~~~
 \delta p   =  \{p , W(p,q)\}= -\frac{\partial W(p,q)}{\partial q}
\label{can}
\eq
 where $W(p,q)$ is an arbitrary function. The fact that these
 transformations preserves the area can be easily   checked
 using    the fact that the general  infinitesimal area-preserving
 diffeomorphism takes the form
\bq
\xi^{a} \rightarrow \xi^{a} + \epsilon^{a}(\xi);~~~\partial_{a}\epsilon^{a} = 0
 \eq
 where $\xi^{a} = (q,p)$.
 General solution  of   $\partial_{a}\epsilon^{a} = 0$   is the sum of the
 two terms
\bq
\epsilon^{a}(\xi) = \epsilon^{ab}\partial_{b}W(\xi) + \sum_{i=1}^{b_{1}}
c_{i}{\bf \epsilon}^{a}
\eq
 where  first term  describes infinite number (all possible functions
 $W(\xi)$)  of the local
 co-exact solutions and the second term describes the finite number
(given by the first Betti number $b_{1}$) of the harmonic forms
 on two-dimensional phase space. It is easy to see that diffeomorphisms
 generated by the first term are nothing but canonical transformations
(\ref{can}).
 Let us note that because our phase space is torus there are also two
  (for torus the first Betti number $b_{1} =2$)   global
 translations $P_{a} = -i\partial_{a}$.

  Any function $f$ on the phase space is transformed under the
 canonical transformation generated by  $W$ according to the rule
 $ w f =\delta_{W} f = \{ f, W \}$, where $w$ is the operator
 corresponding to function $W(\xi)$.
  Using the Jacobi identity
 $\{\{f,W_{1}\} W_{2}\} - \{\{f, W_{2}\} W_{1}\} + \{\{W_{1},W_{2}\}f\}=0$
 one can check that
  algebra of the  area-preserving diffeomorphisms is given
 by the Poisson brackets
\bq
 [w_{1}, w_{2}] f  =
  [\delta_{W_{1}},\delta_{W_{2}}]f = \delta_{\{ W_{1},W_{2}\} } f
\eq
 Let us define our torus phase space $T^{2}$ to be a square with both
 sides equal to $2\pi$ (generalization for the general case is straightforward
 and will be done in the end).
 Any function $W$ can be written in terms of the complete set of harmonics
\bq
W_{\vec{n}} = exp (i \vec{n}\vec{\xi})
\label{W}
\eq
where $\vec{n} = (n_{1}, n_{2})$ with integers $n_{1},n_{2}$
 and $\vec{\xi} = (\xi_{1},\xi_{2})$ are coordinates on the torus.
 One gets the commutation relations for operators $w_{\vec{n}}$
 computing the  Poisson bracket for $W_{\vec{n}}$, \cite{membrane}
  \bq
[w_{\vec{m}}, w_{\vec{n}}] = (\vec{m}\times\vec{n})~ w_{\vec{m} + \vec{n}}
\label{w1}
\eq
where $\vec{a}\times\vec{b} = a_{1}b_{2} - a_{2}b_{1}$.
 One can see that  (\ref{w1}) is
 equivalent to the the commutation
 relations for the $w_{\infty} $ algebra (\ref{w})
 identifying $w_{\vec{m}} = w_{m_{1},m_{2}}$ with
$w^{(m_{2}+1)}_{m_{1}}$.
\footnote{Let us note that this  algebra has a central extension (which does
  not exist
 in the case of the area preserving diffeomorphisms on the sphere)
\cite{membrane}
$[w_{\vec{m}},w_{\vec{n}}] = (\vec{m}\times\vec{n})~w_{\vec{m}+\vec{n}}
 +  \vec{a}\vec{m}\delta_{\vec{m} + \vec{n},0}$ }

 Let us note that $w_{\infty}$ algebra for planar electrons in the
 magnetic field was discussed recently in \cite{ctz} as dynamical symmetry
 for the Quantum Hall (QH) system. Formally this is the same Landau
 problem ( the only  difference is that in \cite{ctz}
  phase space was a plane, however one can consider the QH system on  a
 torus).  However physically it is complete different problems -
 in our case we are dealing with gauge field configurations -
 in the QH case - with the many-body fermion system.

 After quantization we get instead of (\ref{W}) the quantum version
 \bq
{\cal W}_{n,\bar{n}} =
:\exp(\frac{2\pi}{k}(n b^{+} - \bar{n} b)): ~
  =
\exp(\frac{2\pi}{k} n b^{+}) exp(-\frac{2\pi}{k} \bar{n} b);
\nonumber \\
 \bigl[ {\cal W}_{n,\bar{n}},{\cal W}_{m,\bar{m}}  \bigr] =
 \bigl(\exp(-\frac{2\pi}{k} \bar{n} m)
 - \exp(-\frac{2\pi}{k}n \bar{m})\bigr){\cal W}_{n+m,\overline{n+m}} = \\
 -2i\sin\frac{2\pi}{k}(n_1m_2 - n_2 m_1)\exp (-\frac{2\pi}{k}(n_1 m_1 +
 n_2 m_2)) {\cal W}_{n+m,\overline{n+m}} \nonumber
\label{W1}
\eq
 Here $n(\bar{n}) = n_{1}\pm  i n_{2}$ where $n_1,~n_2$ are integers and
 the classical limit corresponds to $k \rightarrow \infty$. Let us also
 note that  after rescaling
 \bq {\cal W}_{n,\bar{n}}
=  \exp ( \frac{\pi}{k}
(n_1^2 + n_2^2))\tilde{\cal W}_{n,\bar{n}}
\label{normalization}
\eq
 one can recover from (\ref{W1}) the Fairlie-Fletcher-Zachos (FFZ) algebra
 ( see  D.B. Fairlie et al in \cite{membrane})
 \bq
 \bigl[ \tilde{\cal W}_{n,\bar{n}},\tilde{\cal W}_{m,\bar{m}}  \bigr] =
  -2i \sin\frac{2\pi}{k}(n_1m_2 - n_2 m_1)
\tilde{\cal  W}_{n+m,\overline{n+m}}
\label{tildeW1}
\eq

 It is also interesting to know how this algebra will act on the
 first Landau level, which is the  quantum analog of the classical phase space.
 To define this action one  need to know the wave functions on the first
 Landau level for torus which was studied in \cite{dnhr}.  One starts
 from the general  wave function on the first Landau level for
 Hamiltonian (\ref{H})
\bq
\Psi({\bf A}_1,{\bf A}_2) =
\exp \bigl( -\frac{ik}{8\pi}{\bf A}_1{\bf A}_2+
\frac{ik}{4\pi} p {\bf A}_1
 - \frac{k}{8\pi} ({\bf A}_2 - p)^2  \bigr) = \nonumber \\
 \exp\bigl( -\frac{k}{16\pi} {\bf A}_{z} {\bf A}_{\bar{z}} \bigr)
 \exp\bigl( -\frac{k}{8\pi} p^2 +\frac{k}{16\pi} {\bf A}^{2}_{\bar{z}} +
 i\frac{k p}{ 4\pi} {\bf A}_{\bar{z}} \bigr)
\label{psi}
\eq
 where $p$ is some constant - ``momentum'' in ${\bf A}_{1}$
 direction. Let us note that except the first factor
the wave function depends only on holomorphic combination
 ${\bf A}_{\bar{z}} = A_1 - i A_2$. This can be  obtained from
 the fact that any wave function on first Landau level is annihilated
 by the operator $a$ ( see (\ref{a}))
\bq
 \frac{\partial}{\partial  {\bf A}_z} \Psi({\bf A}_1,{\bf A}_2) +
  \frac{k}{16\pi} {\bf A}_{\bar{z}} \Psi({\bf A}_1,{\bf A}_2) = 0;
\nonumber \\
\Psi({\bf A}_1,{\bf A}_2) =
\exp\bigl(-\frac{k}{16\pi}{\bf A}_z{\bf A}_{\bar{z}}\bigr)
\Phi({\bf A}_{\bar{z}})
\label{Phi}
\eq

 To get the correct wave functions  for simplest case
 of $\tau = i$ (see (16)) from (\ref{psi}) one have to
 sum over all $p =  4\pi n /k$ . It is easy to see that for even
 $k$ there are $k/2$ different classes $p = 4\pi r/k + 2\pi n$,~
 $r = 1,\cdots k/2;~ n \in Z$ which gives $k/2$ basic wave functions.
 \footnote{In general rational case $k = 2m/n$ it will be $mn$
 basis vectors, but here we shall not consider the general case}.
Thus the basic wave functions can be chosen as

\bq
\Psi({\bf A}_1,{\bf A}_2)_{r} =
  \exp(-\frac{k}{16\pi} {\bf A}_{z} {\bf A}_{\bar{z}})
 \exp(\frac{k}{16\pi} {\bf A}^{2}_{\bar{z}}) \times \nonumber \\
\sum_{n}
 \exp [-\frac{\pi k}{2} (n + 2r/k)^{2}  +
i \frac{k {\bf A}_{\bar{z}}}{2}(n + 2r/k)] = \\
\exp(-\frac{k}{16\pi} {\bf A}_{z} {\bf A}_{\bar{z}})
 \exp(\frac{k}{16\pi } {\bf A}^{2}_{\bar{z}})~~
 \theta\left[ \begin{array}{c} 2r/k \\
 0\end{array} \right] \bigl(\frac{ k {\bf A}_{\bar{z}}}
{4\pi}|\frac{i k }{2})  \nonumber
\label{psi1}
  \eq
where the theta functions  are defined as
\bq
\theta\left[ \begin{array}{c} \alpha \\
 \beta \end{array} \right] \bigl(z|\tau) =
 \sum_{n}
 \exp[ i\pi\tau (n+\alpha)^{2} +
 2\pi i (n+\alpha)(z+\beta) ]
\label{theta}
\eq

 Now let us consider the action of the generators (36)
 on the wave functions  (41). From (\ref{definitionb}) and
 (\ref{Phi}) one can see that
\bq
b~\Psi({\bf A}_1,{\bf A}_2) = \exp\bigl(-\frac{k}{16\pi}{\bf A}_z
{\bf A}_{\bar{z}}\bigr)
(-2i \frac{\partial}
{\partial {\bf A}_{\bar{z}}})
\Phi({\bf A}_{\bar{z}}) \nonumber \\
b^{+}~\Psi({\bf A}_1,{\bf A}_2) = \exp\bigl(-\frac{k}{16\pi}{\bf A}_z
{\bf A}_{\bar{z}}\bigr)
(\frac{i k}
{4 \pi} {\bf A}_{\bar{z}})
\Phi({\bf A}_{\bar{z}})
\eq
and then  ${\cal W}_{n,\bar{n}}$ effectively acts only on holomorphic
 factors $\Phi_{r}({\bf A}_{\bar{z}})$:
 \bq
{\cal W}_{n,\bar{n}}\Phi_{r}({\bf A}_{\bar{z}}) =
\exp(\frac{i}{2} n {\bf A}_{\bar{z}}) exp(\frac{4\pi i}{k} \bar{n}
\frac{\partial}{\partial {\bf A}_{\bar{z}}} )\Phi_{r}({\bf A}_{\bar{z}}) =
\nonumber \\
\exp(\frac{i}{2} n {\bf A}_{\bar{z}}) \Phi_{r}({\bf A}_{\bar{z}}
 +  \frac{4\pi i}{k} \bar{n})
\eq
Using (41)  we get after some calculation ($n(\bar{n}) = n_1 \pm n_2$)
\bq
{\cal W}_{n,\bar{n}}\Phi_{r}({\bf A}_{\bar{z}}) =
 \exp\bigl(\frac{\pi}{k}(n_{1}^{2}+n_{2}^{2})\bigr)
 \exp\bigl(\frac{2\pi i}{k}n_{2}(n_{1} + 2r)\bigr)
\Phi_{r+n_{1}}({\bf A}_{\bar{z}})
\label{representation}
\eq
 It is amusing that to get unitary representation, i.e. to preserve
 the normalization of the wave functions $\Phi_{r}({\bf A}_{\bar{z}})$
we  must rescale ${\cal W}_{n,\bar{n}}$ according to
(\ref{normalization}) - then the nonunitary exponential factor
$\exp\bigl(\frac{\pi}{k}(n_{1}^{2}+n_{2}^{2})\bigr)$ disappears.

  Thus we demonstrated that the ground state (first Landau
 level) wave functions in the topologically massive
 gauge theory (not only in the pure Chern-Simons case
 with infinite mass gap) form a unitary representation
  of the FFZ algebra
 (\ref{tildeW1}). Let us note that states at higher Landau levels
 which can be obtained from our ground state
  wave functions $\Psi_{r}$ by the
 action of the $a^{+}$ operator (25)
\bq
\Psi^{l}_{r}({\bf A}_1,{\bf A}_2)
 = \frac{(a^{+})^{l}}{\sqrt{l!}}\Psi_{r}({\bf A}_1,{\bf A}_2)
\eq
 form unitary equivalent representations because generators
 $\tilde{\cal W}_{n,\bar{n}}$ are built from $b$ and $b^{+}$
 operators only and thus commute with $a$ and $a^{+}$.

 It is interesting to note that one can construct the second
$W$ algebra from $a,a^{+}$ operators in the complete analogy with
(36) - then one gets the direct products of two
 $W_{\infty}$ algebras: $W_{\infty}^{a} \otimes W_{\infty}^{b}$.
 The second (``b'') algebra acts on each Landau level, the first
 (``a'') algebra mixes the level - it is  the coherent states
 $|\alpha>_{r} \sim \exp{(\alpha a^{+})}|0>_{r}$ which are transformed in a
 simple way under the action of $W_{\infty}^{a}$ generators.

  Finally let us write some formulas in the
   case of  a general modular parameter $\tau$.  Using  (16) and
 (17) one get instead
 of (36)
 \bq
{\cal W}_{n,\bar{n}} =
\exp(\frac{2\pi Im \tau}{k} n b^{+}) exp(-\frac{2\pi Im \tau}{k} \bar{n} b)
\eq
 where $n = (-i\tau n_{1} + i n_{2})/Im \tau$,~ $\bar{n} =
 (i\bar{\tau} n_{1} - i n_{2})/Im \tau$ and $[b,b^{+}] = k/2\pi Im\tau$.
 The commutator equals to
\bq
\bigl[ {\cal W}_{n,\bar{n}},{\cal W}_{m,\bar{m}}  \bigr] =
\exp \{-\frac{2\pi}{k Im\tau}\bigl(|\tau|^{2} n_1 m_1 +
 n_2 m_2 - Re\tau(n_{2}m_{1}+n_{1}m_{2})\bigr)\}\times \nonumber \\
 -2i\sin\frac{2\pi}{k}(n_1 m_2 - n_2 m_1) {\cal W}_{n+m,\overline{n+m}}
 ~~~~~~~~~~~~~~~~~~~~~~~~~~
\label{tauW}
\eq
and after renormalization  ${\cal W}_{n,\bar{n}}
=  \exp ( \pi|n_{2} - \tau n_{1}|^{2}/k Im\tau)
\tilde{\cal W}_{n,\bar{n}}$ one gets
 the same  FFZ algebra (\ref{tildeW1}).

 The basic
 wave functions are now ($ r = 1, \cdots k/2$)
\bq
\Psi_{r}({\bf A}_1,{\bf A}_2) =
  \exp(-\frac{k}{16\pi Im\tau} {\bf A}_{z} {\bf A}_{\bar{z}})
 \exp(\frac{k}{16\pi Im\tau} {\bf A}^{2}_{\bar{z}}) \times \nonumber \\
\sum_{n}
 \exp[ i\frac{\pi k \tau}{2} (n + 2r/k)^{2}  +
 i\frac{k {\bf A}_{\bar{z}}}{2}(n + 2r/k) ] =  \\
\exp(-\frac{k}{16\pi} {\bf A}_{z} {\bf A}_{\bar{z}})
 \exp(\frac{k}{16\pi Im\tau} {\bf A}^{2}_{\bar{z}})~~
 \theta\left[ \begin{array}{c} 2r/k \\
 0\end{array} \right] \bigl(\frac{ k {\bf A}_{\bar{z}}}
{4\pi}|\frac{k \tau}{2}) \nonumber
\label{psitau}
\eq
where ${\bf A}_{z}= {\bf A}_{1} + \tau {\bf A}_{2},~
{\bf A}_{\bar{z}}= {\bf A}_{1} + \bar{\tau} {\bf A}_{2}$
 and  the theta-functions were defined in (\ref{theta}).

 One can check that generators (\ref{tauW}) act on  wave functions
 (\ref{psitau}) as
$${\cal W}_{n,\bar{n}}\Phi_{r}({\bf A}_{\bar{z}}) =
\exp ( \pi|n_{2} - \tau n_{1}|^{2}/k Im\tau)\exp(i\phi)
\Phi_{r+n_{1}}({\bf A}_{\bar{z}})\nonumber $$  - in a complete
 agreement with (\ref{representation}). After rescaling
 to the $\tilde{\cal W}_{n,\bar{n}}$ the exponential
 factor $\exp ( \pi|n_{2} - \tau n_{1}|^{2}/k Im\tau)$
 disappears and we again get unitary representation.
 Let us also note that
 these wave functions
 ( to be more precise the  h olomorphic parts)
 can be also obtained in the geometric quantization of the pure
 Chern-Simons theory  \cite{witten},\cite{confblo} and give
  the  conformal blocks for the $c=1$ conformal field theory
(for more details see \cite{confblo}). Let us note that even
 in a case of a {\it finite} mass gap $ M = k \gamma/4\pi$
 one gets the $same$ ground state (first Landau level) wave
 functions.

\section{Conclusion}

 We discussed in this letter the  classical and quantum
 canonical symmetry in the $2+1$ gauge theory with  a
 Chern-Simons term and found  that one can consider
 canonical transformation  on the reduced
 Chern-Simons phase  space  (which becomes after
 the quantization the first Landau level) as
 the $w_{\infty}$ algebra, which after quantization
 becomes FFZ  $W_{\infty}$ algebra (\ref{tildeW1}). This algebra
 (we shall call it $W_{\infty}^{b}$ algebra) commute with the
  Hamiltonian  ${\cal H} \sim (a a^{+} + a^{+} a)$ and thus
 acts independently on each Landau level.
The ground state
 wave functions form the unitary representation
 of this algebra - as well  as  wave functions
 at any Landay level. One can construct another
$W$ algebra from $a,a^{+}$  operators and
  gets: $W_{\infty}^{a} \otimes W_{\infty}^{b}$.
 The second (``b'') algebra acts on each Landau level, the first
 (``a'') algebra mixes the level and acts in a simple form
onto the coherent states
 $|\alpha>_{r} \sim \exp{(\alpha a^{+})}|0>_{r}$.

It is interesting to find if there any connection between
 this algebra and $W_{\infty}$ arising in the $c=1$ strings
 and corresponding matrix models \cite{kpw}-\cite{matrix}.
Let also note that one can get $W_{N}$-algebra from
 the $SU(N)$ Chern-Simons theory \cite{bfk} - thus one can
 get  $W_{\infty}$ for the $SU(\infty)$ Chern-Simons theory.
It is  interesting and open question if there is any
 connection
 between these two different  $W_{\infty}$  structures:
 the first one which is connected with the canonical transformations
 - area preserving  diffeomorphisms
 acting in the phase space (or Hilbert space after quantization)
 and the second one which was considered in \cite{bfk} and  is
 connected with the general coordinate transformations in the
 usual space ( $2+1$ space-time).

{\bf Acknowledgments}

 It is a pleasure to  thank G. Semenoff and  N. Weiss
 for interesting  discussions and  hospitality at
  the University of British
Columbia. I would like to  thank D. Eliezer, A. Morozov and
  C. Trugenberger  for discussions.

\end{document}